\documentclass[a4paper]{spie}   

\usepackage[dvipsnames]{xcolor}
\usepackage{amsmath,amsfonts,amssymb}
\usepackage{graphicx}
\usepackage{xparse} 
\usepackage{lastpage}
\usepackage[colorlinks=true, allcolors=blue]{hyperref}

\usepackage{fancyhdr}
\renewcommand{\headheight}{25pt} 
\renewcommand{\headsep}{40pt} 

\NewDocumentCommand{\revisionLM}{m}{{#1}}

\NewDocumentCommand{\revisionAF}{m}{{#1}}

\NewDocumentCommand{\revisionPTP}{m}{{#1}}
\NewDocumentCommand{\revisionFC}{m}{{#1}}

\graphicspath{{figures/}}

\title{Co-design approach to aperture masking for imaging through atmospheric turbulence}

\author[a]{Olivier Herscovici-Schiller}
\author[b]{Laurent M. Mugnier}
\author[c]{Alice Fontbonne}
\author[a]{Pauline Trouv\'e-Peloux}
\author[a]{Fr\'ed\'eric Champagnat}
\author[b]{Yann Lucas}
\author[b]{Yann Lai-Tim}
\author[a]{Fr\'ed\'eric Cassaing}
\affil[a]{DTIS, ONERA, Université Paris-Saclay, 91120 Palaiseau, France}
\affil[b]{DOTA, ONERA, Université Paris-Saclay, 92320 Châtillon, France}
\affil[c]{DOTA, ONERA, Université Paris-Saclay, 91120 Palaiseau, France}

\authorinfo{firstname.lastname@onera.fr, with firstname.lastname $\in$ \{olivier.herscovici, laurent.mugnier, alice.fontbonne, pauline.trouve, frederic.champagnat\} }

\begin{document}
\maketitle

\begin{abstract}
Aperture masking interferometry is a technique originally designed to alleviate the influence of atmospheric turbulence on images recorded on ground-based telescopes. In this communication, we explore the optimization of the aperture mask by an optical/digital co-design approach in order to obtain diffraction-limited images of relatively bright objects imaged through turbulence. We show that, with a few simplifying assumptions, it is possible to express the Mean Square Error of the restored image as a function of the chosen mask, of the spatial Power Spectral Density of the observed object and of the noise level, without actually computing any image. This allows us to optimize the aperture mask with a reduced computing cost. We also implement a multi-frame \revisionFC{myopic} 
algorithm to estimate jointly the observed object,  \revisionFC{the piston and the tip-tilt} in front of each sub-aperture, and check by simulations that the aperture masks obtained indeed allow a satisfactory image reconstruction.
\end{abstract}

\section*{INTRODUCTION}
\label{sec:intro}

The angular resolution of ground-based optical imaging is severely degraded by atmospheric turbulence. Adaptive optics (AO) is the reference technique~\cite{Roddier99} to compensate this effect. However, adaptive optics is relatively complex and costly. In this work, we propose a\revisionAF{n optical/digital} co-design approach \revisionAF{to restore angular resolution.} 

Without adaptive optics, the high spatial frequencies of a long-exposure image are irreversibly lost, whereas short-exposure images retain information up to the diffraction cutoff frequency of the telescope~\cite{Roddier81,Fried66}. 
Interpreting image formation as an interferometric process~\cite{Mariotti89} led Rhodes and Goodman~\cite{Rhodes73} to propose aperture masking as a way to avoid the loss of high spatial frequencies in turbulent short-exposure images. By suppressing redundancy \revisionLM{(defined below)} in the pupil, the technique limits the impact of turbulence on the Optical Transfer Function (OTF) to a phase term, without attenuating its modulus, hence without lowering the signal-to-noise ratio (SNR). 

Aperture masking relies on the interpretation of image formation as the superposition of interference patterns produced by all pairs of points in the entrance pupil. \revisionAF{At wavelength $\lambda$, each }
measured spatial frequency $u_0$ originates from the pairs of sub-apertures separated by $\lambda u_0$,  \revisionAF{because each pair produces }
a sinusoidal fringe pattern in the focal plane. If a given spatial frequency is created by a single pair of sub-apertures, then the effect of aberrations is merely a shift of the corresponding fringes, i.e. a phase term in the OTF, without any reduction of the Modulation Transfer Function (MTF). When \revisionLM{the pupil is redundant, \emph{i.e.}, when} several pairs share the same separation and orientation, turbulence de-phases the corresponding fringe patterns with respect to one another; they blur mutually and the MTF drops.

A non-redundant aperture mask is therefore designed to cover all spatial frequencies of interest while avoiding redundancy. The technique is now known under several names: Aperture Masking Interferometry, Sparse Aperture Masking, Non-Redundant Masking~\cite{Baldwin86,Haniff87,Lacour11,Ramos24}. 

\section{Co-design Strategy and Performance Criterion}
\label{sec:codesign}

The imaging system is conceptually divided in two parts: an optical part and a processing part. The optical part is the telescope with a masked aperture. Its function is to acquires stacks of short-exposure images of the object.
The processing part is a computer with a restoration algorithm. Its function is to take the stacks as input and deliver an estimate of the object as output. The optical part and the processing part are designed jointly. The goal is to minimize the mean square error (MSE) between the observed object and the object restored from the image stack.

\begin{figure}[htb]
\begin{center}
\includegraphics[width=0.95\textwidth]{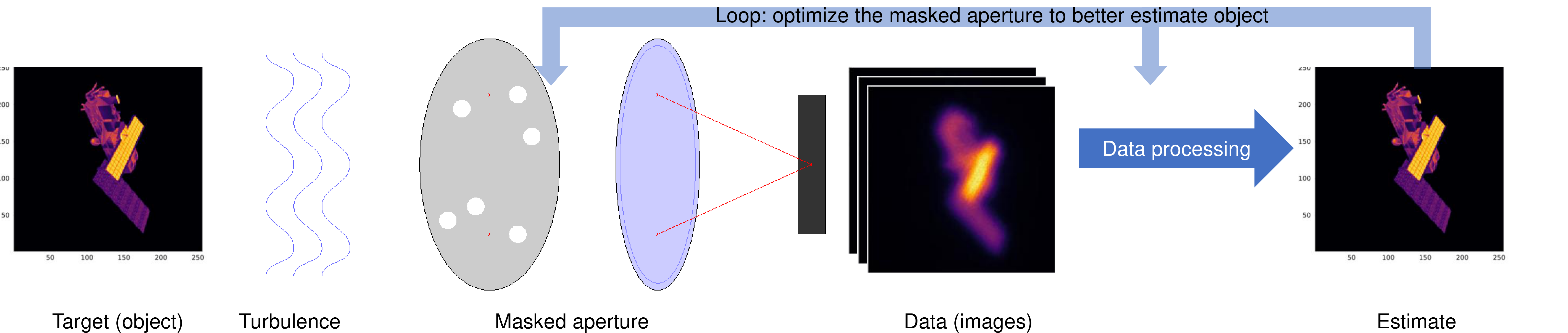}
\end{center}
\caption[flow]{\label{fig:flow}
Co-design of the masked aperture and of the post-processing algorithm. The data stacks are short-exposure images acquired through turbulence and the aperture. The object estimate is obtained by myopic deconvolution. The mask geometry is optimized in an outer loop.}
\end{figure}

\section{A tractable MSE criterion}

\revisionPTP{A tractable MSE criterion can be derived considering }
a multi-frame Wiener filter for $K$ images, with inputs the turbulent OTFs, the Power Spectral Density (PSD) of the target, and the PSD of the noise. \revisionLM{The} \revisionPTP{theoretical} \revisionLM{MSE can then be expressed, on average for infinitely many images, as a function of the set of OTFs and the two abovementioned PSDs without actually simulating images.}
However, this expression requires drawing turbulent OTFs, which is \revisionLM{yet} too costly for optimization in practice. Therefore we derived an asymptotic expression \revisionPTP{of the MSE, referred to as MSE$_a$,} for a long-exposure take, under the hypothesis that the residual phase over each sub-aperture reduces to an independent piston,
\begin{equation}
\label{eq:mse}
\mathrm{MSE}_a = \int_{[0,1]^2}
\frac{S_o(\nu)\,S_n(\nu)}
{K\,S_o(\nu)\left( N_\mathrm{pup}^2\,|\tilde h(\nu)|^2 + \sum_{m\neq n} \left|\tilde h(\nu+\nu_m-\nu_n)\right|^2 \right) + S_n(\nu)}\; \mathrm{d}\nu ,
\end{equation}
where $\tilde h(\nu)$ is the OTF of a single centred sub-aperture (divided by the number of sub-apertures), $\nu_m$ the centre frequency of sub-aperture $m$, $N_\mathrm{pup}$ the number of sub-apertures, $K$ the number of frames, $S_o$ the PSD of the target and $S_n$ the PSD of the noise. This expression depends only on the pupil configuration, on the prior on the PSD of the scene, and on the prior on the PSD of the noise models.
It does not require the simulation of turbulent images, so we can use it as a criterion in the optimisation loop.

\begin{figure}[htb]
\begin{center}
\begin{tabular}{c}
\includegraphics[width=0.7\textwidth]{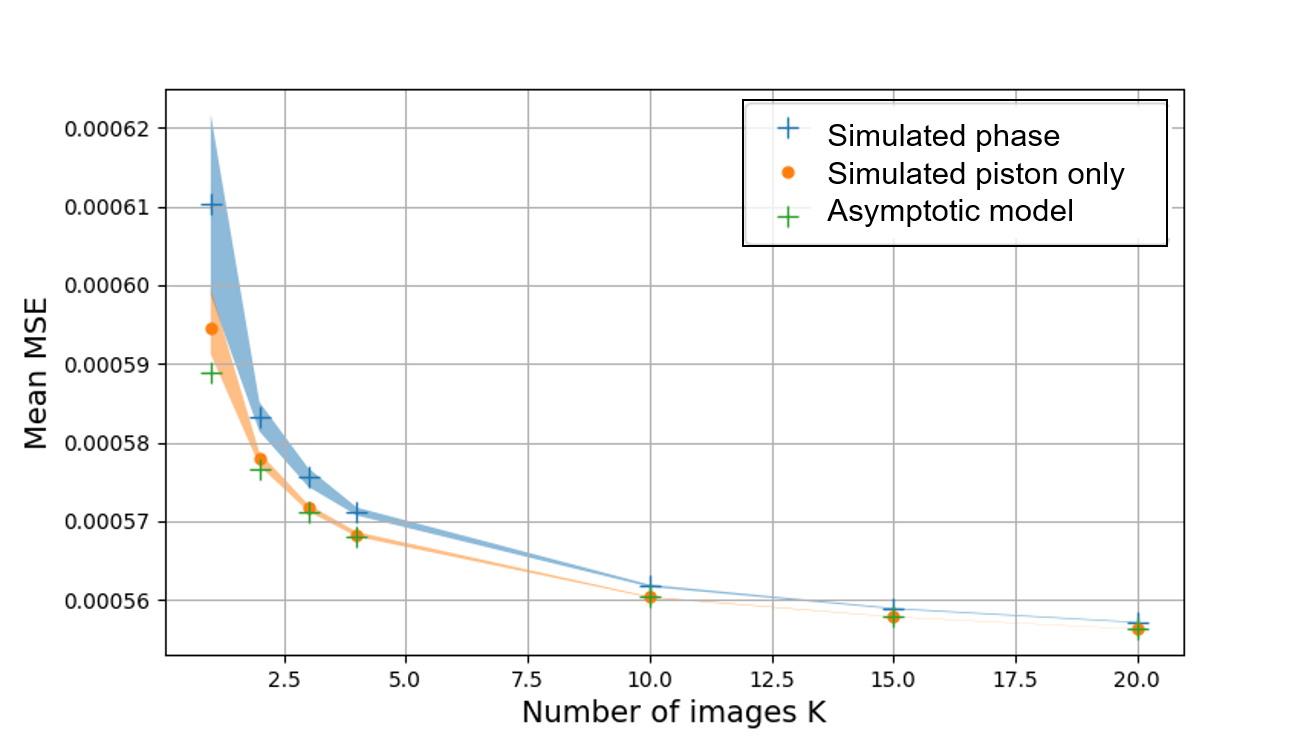}
\end{tabular}
\end{center}
\caption[masks]{\label{fig:varK}
\revisionPTP{Variation of the theoretical MSE corresponding to a multiple images Wiener filter  with the number of images $K$, and the asymptotic MSE$_a$ of equation \ref{eq:mse}. Parameters are  $S_o(\nu)\propto\nu^{-1.5}$,\revisionPTP{$S_n = $}$N_\mathrm{photons}=10^7$, for a Goley-6 configuration.}}
\end{figure}

On Fig~\ref{fig:varK} \revisionAF{}, we plot the asymptotic MSE$_a$ and the theoretical MSE calculated with simulated turbulent Point Spread Functions (PSFs) \revisionPTP{, for a Goley-6~\cite{Golay71,Cassaing18} configuration with  PSD $S_o(\nu)\propto\nu^{-1.5}$,  $S_n=N_\mathrm{photons}=10^7$, assuming Poisson noise only}. We regard the asymptotic limit as valid for $K \gtrsim 10$~frames.

\section{MASK OPTIMISATION}
\label{sec:optim}

For a given PSD of the \revisionPTP{noise} 
and of the object, we find the configuration of the aperture which minimizes the asymptotic \revisionPTP{MSE$_a$}. We fix the total collecting area in order to compare configurations with fixed SNR. \revisionPTP{The optimization is conducted with} 
the Covariance Matrix Adaptation Evolution Strategy (CMA-ES)~\cite{Hansen06}, which handles well non-convex gradient-free problems. 
\revisionFC{We use penalties to constrain 
the sub-apertures to stay inside the overall mask aperture}.

\begin{figure}[htb]
\begin{center}
\begin{tabular}{c}
\includegraphics[width=0.78\textwidth]{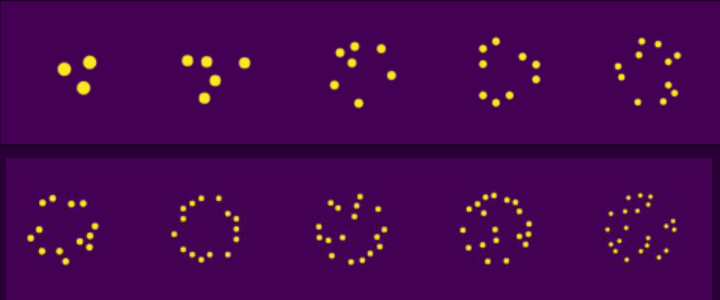}
\end{tabular}
\end{center}
\caption[masks]{\label{fig:masks}
Optimized masks for various numbers of sub-apertures for object PSD $S_o(\nu)\propto\nu^{-1.5}$, $N_\mathrm{photons}=10^7$, with a fixed total collecting area).}
\end{figure}

The optimized masks (Fig.~\ref{fig:masks}) are compact non-redundant arrangements.
For some configurations, like six sub-apertures, we find the classical Golay-6 configuration~\cite{Golay71,Cassaing18}. 

\begin{figure}[htb]
\begin{center}
\begin{tabular}{c}
\includegraphics[width=0.52\textwidth]{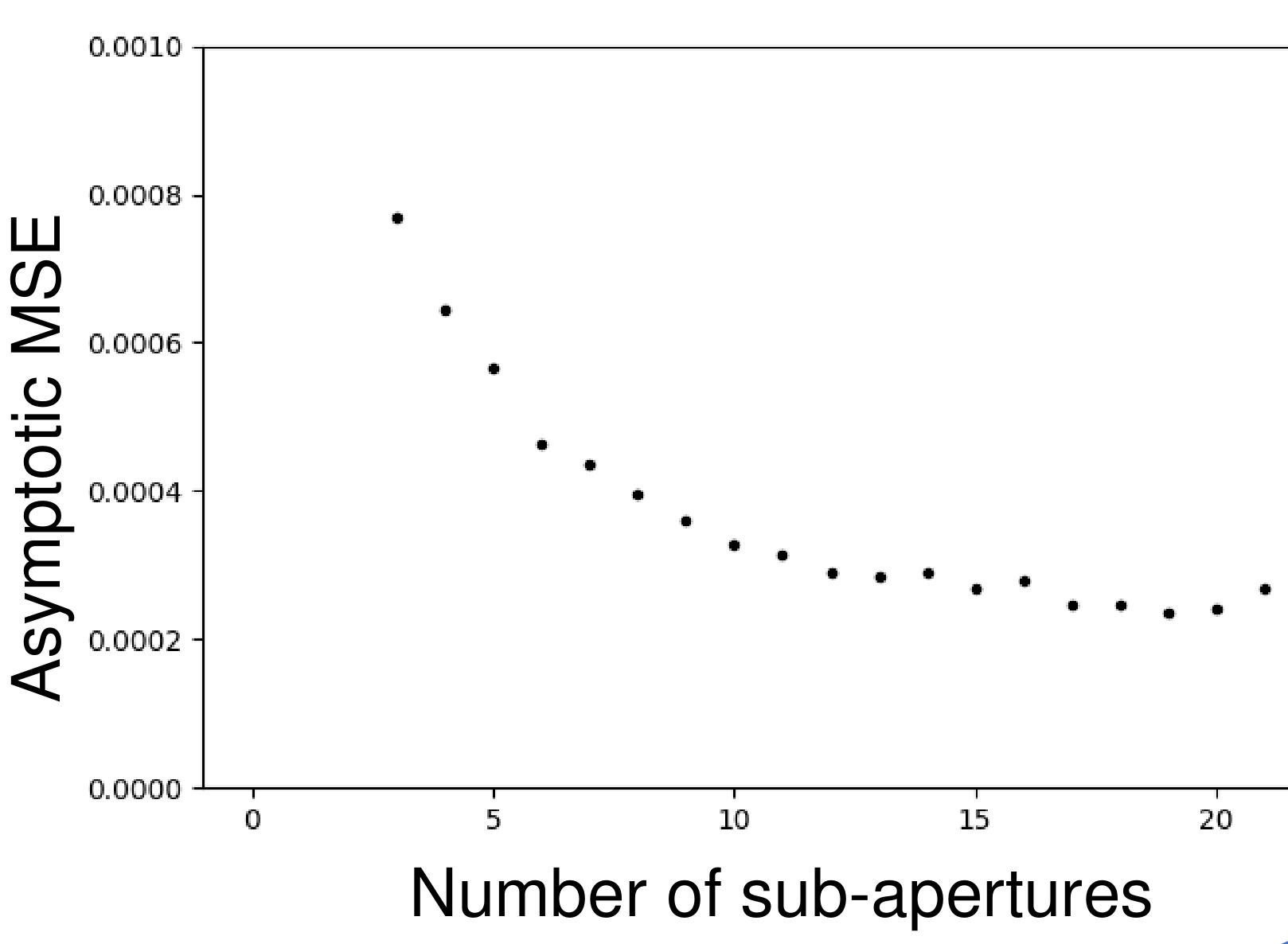}
\end{tabular}
\end{center}
\caption[msen]{\label{fig:msen}
Asymptotic MSE as a function of the number of sub-apertures ($S_o(\nu)\propto\nu^{-1.5}$, $N_\mathrm{photons}=10^7$, total collecting area constant).}
\end{figure}

In Fig.~\ref{fig:msen}, we display the asymptotic MSE as a function of the number of sub-apertures. The error decreases markedly in the beginning, then hits a floor between 10 and 15 sub-apertures.
Since the number of turbulent parameters to be estimated by the myopic deconvolution grows with $N_\mathrm{pup}$, we choose to take one of the smallest reasonable numbers of sub-apertures, which we decide to be eleven.

\section{Myopic Image Reconstruction}
\label{sec:myopic}

In this section, we fix the mask. We have a stack of simulated data, $y_t$, obtained by passing $o$ (which we assume to be the same for the whole stack) through the instantaneous turbulent PSF $h_t$ plus a Gaussian noise: 
\begin{equation}
y_t = h_t \star o + b_t .
\end{equation}

We parameterize $h_t$ by the first Zernike modes over each sub-aperture~\cite{Mugnier06}: piston plus tip-tilt. 
We can then jointly estimate $o$ and the parameters of $h_t$ for each sub-aperture and each frame as~\cite{Mugnier01}
\begin{equation}
\label{eq:myopic} 
(\hat o, \hat a) = \arg\min_{o\ge 0,\ a}\ \sum_{t} \left\| h_t(a) \star o - y_t \right\|^2 + \mu \int \|\nabla o\| \quad \mathrm{under\; constraint} \quad o\ge0.
\end{equation}
The first term is a likelihood term which measures a distance between the model and the actual data; the second term is a total-variation regularization which\revisionLM{, in practice,} imposes a sparse well-defined boundary to the object. The constraint that light emitted be positive helps the optimizer. In practice, we use L-BFGS-B to minimize the criterion.

\begin{figure}[htb]
\begin{center}
\begin{tabular}{c}
\includegraphics[width=0.38\textwidth]{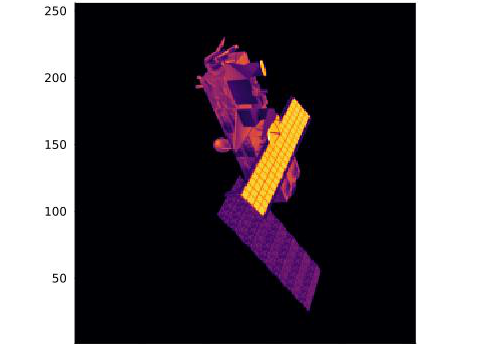}
\end{tabular}
\end{center}
\caption[target]{\label{fig:target}
Object in the end-to-end simulations.}
\end{figure}

The target (Fig.~\ref{fig:target}) is a realistic satellite whose optical signature was simulated by ONERA with a physics-based tool~\cite{hoarau2019interactive,Coiro24,mercier2025theoretical}. We generated stacks of $T=20$ short-exposure images for a turbulence level corresponding to $D/r_0=7$ using a full end-to-end simulator (isoplanetic, von Kármán turbulence), so that we can assess the robustness of our inversion to a model error.

\begin{figure}[htb]
\begin{center}
\begin{tabular}{c}
\includegraphics[width=0.86\textwidth]{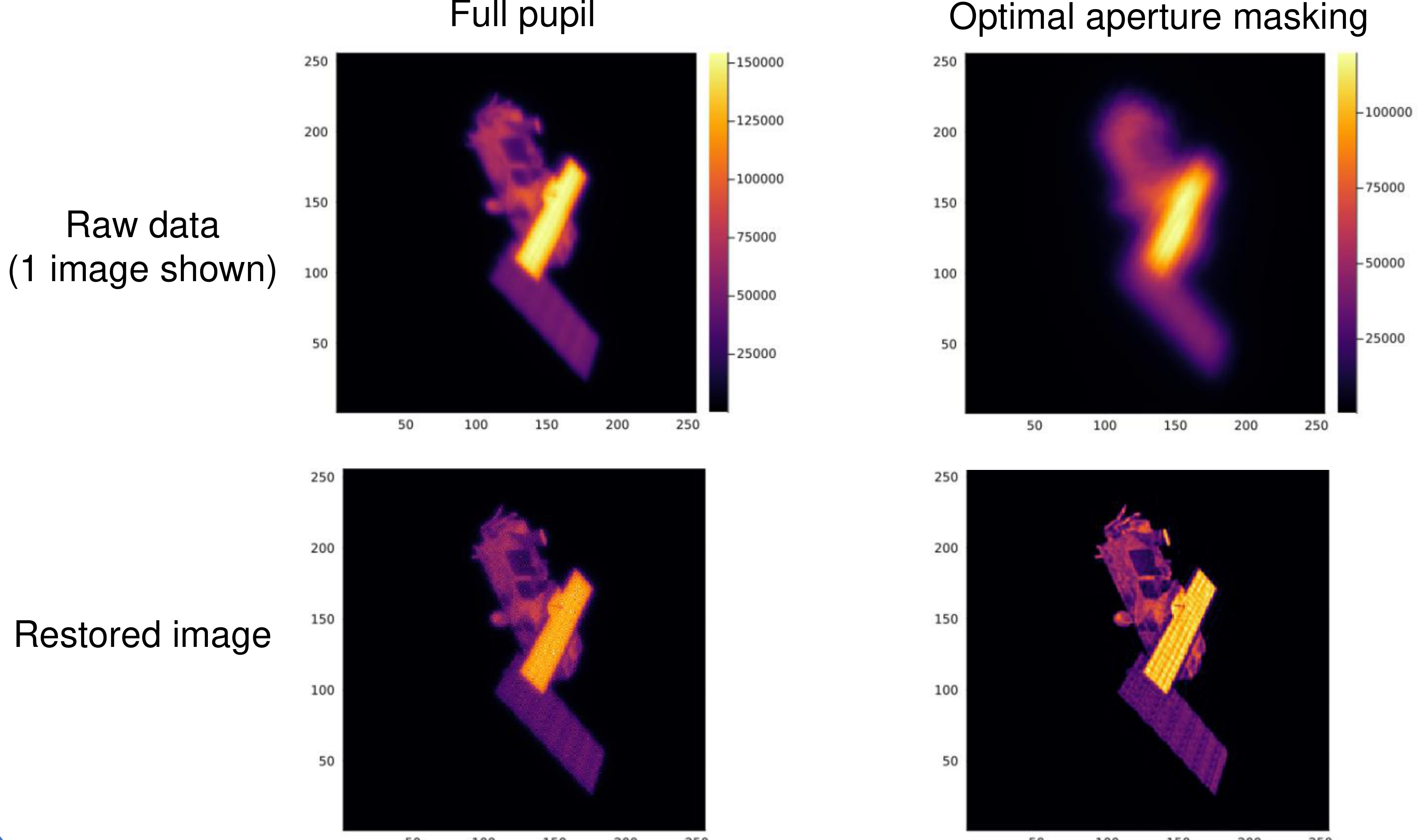}
\end{tabular}
\end{center}
\caption[recon]{\label{fig:recon}
Full pupil (left) versus optimized aperture masking (right). Top: raw data (one frame). Bottom: object restored by myopic deconvolution.}
\end{figure}

On Fig.~\ref{fig:recon}, we show in simulation the data and reconstruction of our proposed device and, for comparison, a full-pupil telescope whose diameter equals the largest sub-aperture separation of the mask, so that both instruments reach the same spatial frequencies. Furthermore, \revisionLM{in order to make a fair comparison, in our simulation} the full pupil receives \revisionLM{significantly} more photons. The better reconstruction quality \revisionLM{obtained with the masked aperture} indicates that aperture masking is a convenient way to mitigate turbulence in situations where AO would be too complicated or costly.

\section{CONCLUSION AND PERSPECTIVES}
\label{sec:conclusion}

In this proceeding, \revisionAF{we showed an optical-digital co-design approach to optimize aperture masking as } 
a way to mitigate turbulence in situations where AO would be too complicated or costly.
We introduced an asymptotic performance criterion which is cost-efficient enough to be used in a loop of optimisation, \revisionLM{and} we used it to optimise the aperture. \revisionLM{We also implemented a myopic multi-frame image deconvolution algorithm}, and we \revisionLM{used it to validate our co-design method in end-to-end simulations.} 

In future work, we will test the robustness to anisoplanatic imaging, we will extend the formalism to a wider spectral band\revisionAF{. We also plan to make an experimental demonstration on an in-lab testbed, and then move to observation data on a telescope.}

\acknowledgments
\revisionLM{The authors thank Ugo Tricoli for providing the satellite scene used in this article.}

\bibliography{aperture_masking}
\bibliographystyle{spiebib}

\end{document}